# Mineralogical Characterization and Phase Angle Study of Two Binary Near-Earth Asteroids, Potential Targets for NASA's Janus Mission

Lucille Le Corre[1,2,3], Juan A. Sanchez[2], Vishnu Reddy[4,3], Adam Battle[4,3], David Cantillo[4,3], Benjamin Sharkey[4,3], Robert Jedicke[5,3], Daniel Scheeres[6,3].

## Abstract

Ground-based characterization of spacecraft targets prior to mission operations is critical to properly plan and execute measurements. Understanding surface properties, like mineralogical composition and phase curves (expected brightness at different viewing geometries) informs data acquisition during the flybys. Binary near-Earth asteroids (NEA) (35107) 1991 VH and (175706) 1996 FG3 were selected as potential targets of the National Aeronautics and Space Administration's (NASA) dual spacecraft Janus mission. We observed 1991 VH using the 3-m NASA Infrared Telescope Facility (IRTF) on Mauna Kea, Hawaii, on July 26, 2008. 1996 FG3 was observed with the IRTF for seven nights during the spring of 2022. Compositional analysis of 1991 VH revealed that this NEA is classified as an Sq-type in the Bus-DeMeo taxonomy classification, with a composition consistent with LL ordinary chondrites. Using thermal modeling, we computed the thermally corrected spectra for 1996 FG3 and the corresponding best fit albedo of about 2-3% for the best spectra averaged for each night. Our spectral analysis indicates that this NEA is a Ch-type. The best possible meteorite analogs for 1996 FG3, based on curve matching, are two carbonaceous chondrites, Y-86789 and Murchison. No rotational variation was detected in the spectra of 1996 FG3, which means there may not be any heterogeneities on the surface of the primary. However, a clear phase reddening effect was observed in our data, confirming findings from previous ground-based studies.

[1] Corresponding author lecorre@psi.edu
[2] Planetary Science Institute, 1700 East Fort Lowell Road, Tucson, AZ 85719, USA
[3] Visiting Astronomer at the Infrared Telescope Facility (IRTF)
[4] Lunar and Planetary Laboratory, University of Arizona, 1629 E University Blvd, Tucson, Arizona 85721, USA
[5] Institute for Astronomy, University of Hawaii, 2680 Woodlawn Drive, Honolulu, HI 96822-1839, USA
[6] Ann and H.J. Smead Department of Aerospace Engineering Sciences, University of Colorado at Boulder, 3775 Discovery Drive, Boulder, CO 80303, USA

1. INTRODUCTION

Binary near-Earth asteroids (NEA) (175706) 1996 FG3 and (35107) 1991 VH are the two potential targets of NASA's SIMPLEX Janus mission. Each of Janus' spacecraft will perform a flyby of one of these targets and take detailed observations of the binary NEA. 1996 FG3 was discovered at Siding Spring Observatory (Australia) on March 24, 1996 and is a potentially hazardous Apollo asteroid, although it currently has negligible probability of an Earth impact in the foreseeable future. The binary nature of this asteroid was discovered by Pravec et al. (2000). 1996 FG3 was classified as a C-type in the second phase of the Small Main-belt Asteroid Spectroscopic Survey (SMASSII) by Bus and Binzel (2002) and has a geometric albedo of 0.045±0.002 (Yu et al. 2014). The primary of this binary system has a rotation period of 3.595 h, a diameter of 1.69±0.22 km (Scheirich et al. 2015), and a density of ~1.4 g/cm$^3$ (Scheirich & Pravec 2009). The secondary has a diameter of 0.49±0.08 km (Scheirich et al. 2015) and a rotation period of 16.15±0.01h (Pravec et al. 2006). Wolters et al. (2011) observed 1996 FG3 in the thermal infrared and inferred that the surface is covered in dust based on the low thermal inertia. This NEA was also observed by several teams in the visible and near-infrared (Binzel et al. 2012, Binzel et al. 2019, de León et al. 2013, Fornasier et al. 2014, Perna et al. 2013, Rivkin et al. 2013) because it was originally the target of the *Marco Polo-R* sample return mission concept proposed to the European Space Agency. Based on these spectra, possible carbonaceous chondrites meteorite analogs were identified such as CM2 (de León et al. 2011, Popescu et al. 2012) and CV3 (Rivkin et al. 2012). These observations also show varying spectral slopes suggesting either surface heterogeneity or observational geometry effects due to different phase angles. Our observational campaign presented in this work is the first step towards resolving this conundrum by obtaining a consistent dataset of near-IR spectra of this NEA during its flyby of the Earth in Spring 2022. By going to 1996 FG3, a primitive body with possible heterogenous surface (de León et al. 2013, Perna et al. 2013, Rivkin et al. 2013), the Janus mission can address these fundamental questions by documenting the relative diversity of distinct regions on this binary asteroid that is likely to be a rubble pile.

One of NASA's Janus mission objectives is to observe 1996 FG3 at multiple phase angles around closest approach, at least three on approach and three on departure, to constrain surface morphology and refine asteroid component shapes to resolve decameter-scale features. The rationale for this objective is that binary asteroid formation theories predict mass movement morphologies at the decameter scale and should leave evidence at a commensurate scale. The observation requirements will support the development of surface morphology measurements and shape refinements of both bodies. The IRTF observations presented in this work will verify if the observed surface heterogeneity on 1996 FG3 is due to surface variations or due to different phase angles.

If rotationally-resolved spectra on one night show variations in spectral slope (increasing or decreasing reflectance as a function of wavelength), it would be an indication that surface heterogeneity is the likely cause. This would suggest that the surface changes from boulders to regolith as the object rotates, and could be confirmed by the cameras on-board the Janus spacecraft during its flyby of 1996 FG3. If the spectral slopes of rotationally-resolved spectra from a single night remain constant (no change with rotation), but the average spectra of the entire rotation phase taken at different phase angles show variations in spectral slope, then phase angle would be the likely cause of the observed spectral variations. The Janus spacecraft observations at various phase angles during its encounter with 1996 FG3 could confirm this. The results from our observation campaign will also help cross-calibrate NASA IRTF point source spectra with resolved observations from a spacecraft, thereby improving our confidence in Earth-based remote characterization of near-Earth asteroids. This is the last opportunity to characterize 1996 FG3 before the Janus mission encounter with the target given that the target will not get as bright after 2022.

1991 VH was also discovered at the Siding Spring Observatory, on November 1, 1991. The binary nature of this asteroid was discovered by Pravec et al. (1998). 1991 VH was classified as a Sk-type in the SMASSII survey and Sq-type in the Bus-DeMeo survey (DeMeo et al. 2009). It has a geometric albedo of 0.27±0.16 (Mueller et al. 2011). The primary in this binary system has a rotation period of 2.6236 h (Pravec et al. 2006), a diameter of ~1.2 km, and a density of ~1.5 g/cm$^3$ (Naidu et al. 2018). This NEA was also studied by Binzel et al. (2019) using visible and near-infrared observations from 2002. No detailed compositional characterization is available in the literature for this object therefore we present in this work the first comprehensive mineralogical analysis.

## 2. OBSERVATIONS AND DATA REDUCTION

Near-infrared (NIR) observations of asteroid 1991 VH were done on July 26, 2008 and for asteroid 1996 FG3, in March and April 2002, with the 3-m NASA Infrared Telescope Facility (IRTF) on Mauna Kea, Hawaii. For both NEAs, the NIR spectra were acquired using the SpeX low-resolution (R ~100) spectrometer (Rayner et al. 2003) in prism mode (0.7–2.5 µm). Since asteroid surfaces are dominated by minerals that produce broad crystal field absorption features, the low-resolution prism mode is ideal for getting high signal to noise ratio (SNR) spectra with the appropriate spectral coverage (0.7–2.5 µm).

For each target asteroid, images were taken for two different slit positions (A-B) following the sequence ABBA. We selected the best local solar-type (G-type) star for each night of observation to be able to remove the effect of the local atmosphere and telluric bands. Observations for these standard stars were scheduled before and after each set of NEA spectra to spatially and temporally sample the local atmospheric conditions. Solar analog star SAO 120107 was observed to adjust the spectral slope from variations due to the use of a non-solar extinction star (solar continuum correction). Solar analog stars are ratioed to the local G-type star which results in a continuum representing the non-solar slope of the local G-type telluric star. The final correction for the solar continuum involves the division of the telluric corrected NEA spectrum by a non-solar continuum derived in the previous step.

For 1996 FG3 observations, a set of ten to fourteen spectra, each using a ~200-second-long integration, were collected for each rotational phase of the asteroid. In addition, a set of ten ~3-second exposures of local G-type star were obtained on either side of the asteroid set. This block of observations is typically 1-hour long, therefore, a set of four such blocks over a four-hour period is adequate for covering the 3.6-hour rotation period of 1996 FG3. The rest of the time was used for instrument setup, solar analog star observations (~2-second exposures with five cycles), focusing the telescope between sets, and collection of calibration frames (arcs and flats).

Observational circumstances for both 1991 VH and 1996 FG3 are detailed in Table 1. Since the targets were brighter than V Mag. 17, *Guidedog* was used for guiding. The ephemeris and rates are available on JPL Horizons that can interface with the IRTF telescope control system. All of our spectra were calibrated using the IDL-based software *Spextool* (Cushing et al. 2004), which includes a processing step for removing the telluric bands (Vacca et al. 2003). A more detailed explanation of the data reduction method is available in Sanchez et al. (2013) and Sanchez et al. (2015).

Our 1996 FG3 observations include rotationally resolved near-IR spectra of 1996 FG3 over several phase angle ranges. The primary has a rotation period of ~3.6 hours and by observing the object over its entire rotation phase it is possible to detect rotational heterogeneity. Observing at multiple phase angles as this NEA comes towards the Earth for its April 2022 flyby will help answer if phase angle is the primary cause of the observed spectral variations. Our observations sample the phase angle space on at least seven occasions (seven half nights of observing) between the minimum phase angle of 12.5 deg and a maximum of 53.2 deg (Table 1). This range covers the observed phase angle range of published data (8-58 deg). Previous data have been collected over multiple epochs by multiple observers using different telescopes and processed with a variety of data reductions protocols. The 2022 apparition of 1996 FG3 provided us with a unique opportunity to collect a consistent dataset with one telescope and instrument and to process all data sets in a self-consistent manner.

**Table 1.** Date and observation circumstances for the IRTF data of both NEAs 1991 VH and 1996 FG3.

| Target | Date (UTC) | UTC time | Phase Angle (Degree) | V mag | Standard Star | Solar Analog | Airmass (Solar Analog-Asteroid)[a] |
|---|---|---|---|---|---|---|---|
| **1991 VH** | 7/26/08 | 05:53–07:41 | 85.7 | 15.1 | SAO 45427 | SAO 120107 | 1.14–1.17 |
| **1996 FG3** | 3/22/22 | 09:30–12:20 | 23.7 | 17.1 | SAO 158158 | SAO 120107 | 1.24–1.44 |
| **1996 FG3** | 4/8/22 | 07:20–13:20 | 12.5 | 15.4 | SAO 181142 | SAO 120107 | 1.13–1.46 |
| **1996 FG3** | 4/15/22 | 04:50–11:50 | 18.7 | 15.1 | HD 106991 | SAO 120107 | 1.07–1.43 |
| **1996 FG3** | 4/20/22 | 04:50–11:20 | 29.9 | 15 | HD 100499 | SAO 120107 | 1.04–1.40 |
| **1996 FG3** | 4/23/22 | 04:50–10:20 | 38.8 | 15 | SAO 156410 | SAO 120107 | 1.07–1.31 |
| **1996 FG3** | 4/25/22 | 04:50–09:00 | 45.6 | 15 | SAO 179085 | SAO 120107 | 1.02–1.31 |
| **1996 FG3** | 4/27/22 | 04:50–09:50 | 53.2 | 15.1 | SAO 155869 | SAO 120107 | 1.03–1.34 |

[a] For each night, the air mass is provided for the solar analog and for the asteroid (average airmass computed from the spectral sets used to calculate the average spectrum for each night).

## 3. RESULTS

The NIR spectrum of 1991 VH is shown in Figure 1. The spectrum has two well-defined absorption bands over the 0.8-2.5-$\mu$m range. The first feature is centered around 1 μm and the second one around 2 μm, consistent with S-complex objects. NIR spectra of 1996 FG3 are shown in Figure 2. Besides the two features seen in some spectra due to the incomplete telluric band correction (centered at ~1.4 and ~1.9 μm), spectra appear essentially featureless in the 0.8–2.5-μm range with the exception of a weak absorption band around 1.13 μm due to phyllosilicates (*e.g.*, Cloutis et al. 2011). Our spectra are generally similar to the ones presented in de León et al. (2013) but do not have the slope change around 1.6 μm visible in their spectra acquired in 2011. Their spectra from 2011 have a red slope from ~ 0.7 to 1.6 μm, and then become less steep from 1.6 to 2.5 μm. Interestingly, a similar slope change at 1.6 μm was also seen in our IRTF observations of Ryugu (Le Corre et al. 2018). The discrepancies between the spectra obtained in previous work could be due to higher humidity, however, this is not specified in their paper. The relative humidity at the time of our observations was elevated (~30-50%) and sometimes variable. Our 1996 FG3 spectra also show a sharp rise in reflectance beyond 2.0 μm due to the thermal tail of the Planck curve (Reddy et al. 2012). This is because 1996 FG3 is a low albedo object, in contrast to 1991 VH, which is a higher albedo S-type asteroid and therefore does not have a thermal tail in the NIR wavelength range (<2.5 μm).

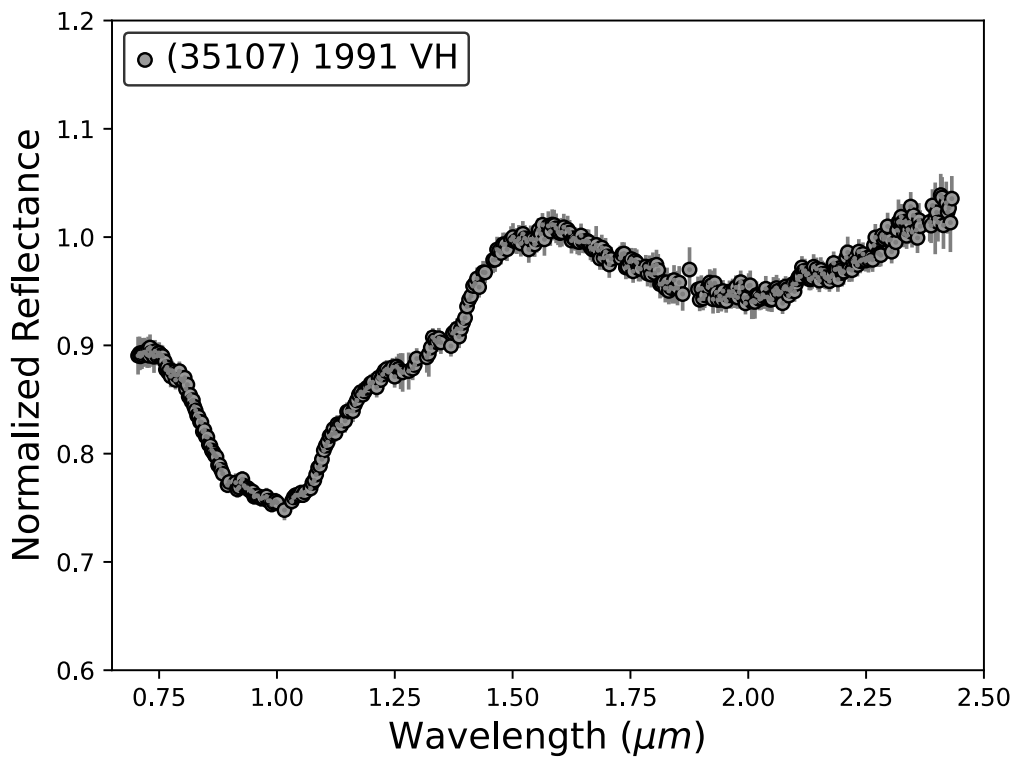

*Figure 1.* NIR spectrum of (35107) 1991 VH (gray circles) normalized to unity at 1.5 microns with prominent absorption bands due to minerals olivine and pyroxene.

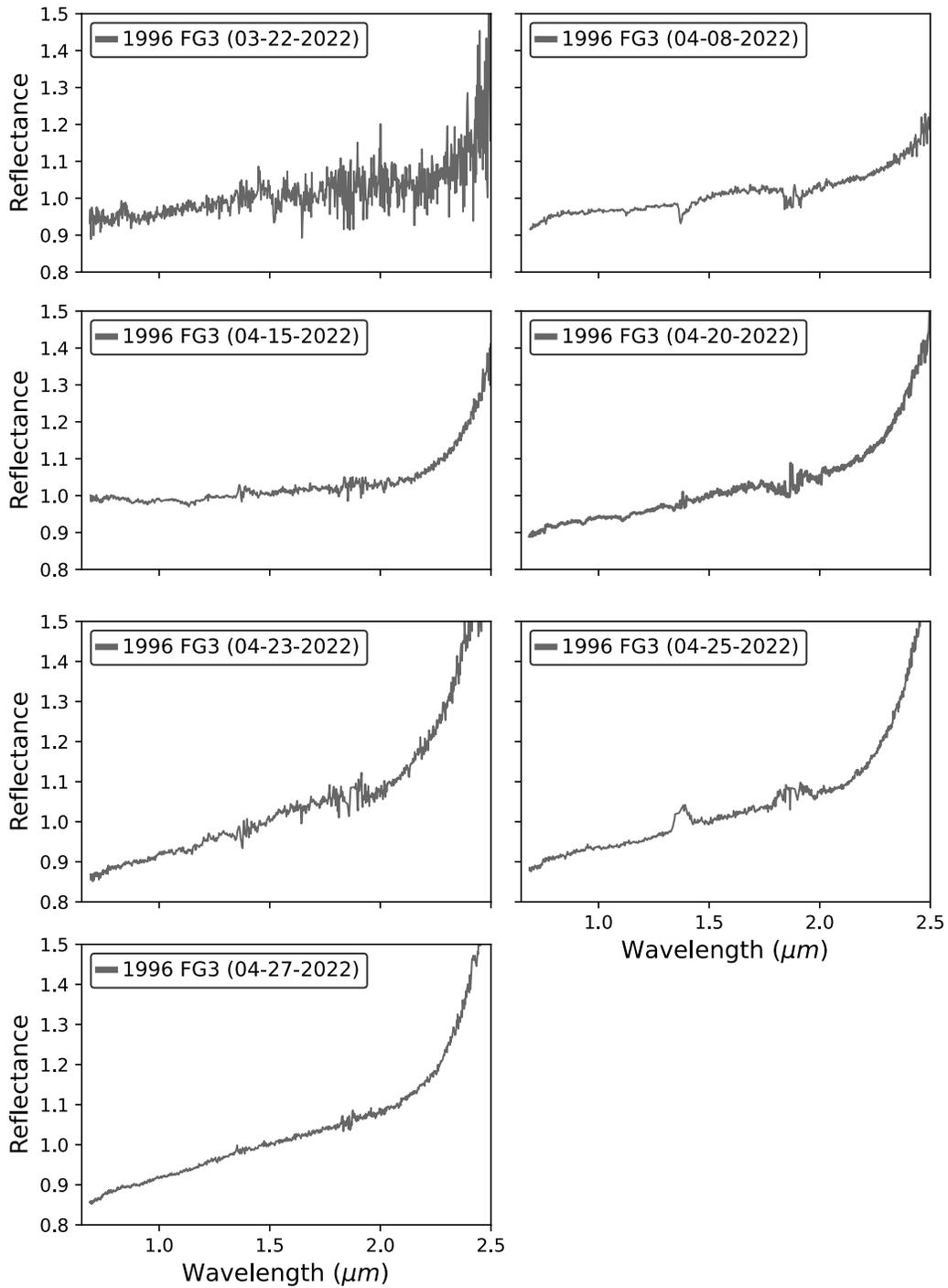

**_Figure 2._** NIR spectrum of (175706) 1996 FG3 for each night of observation. The best sets of spectra (based on SNR, weather conditions and telluric bands residuals) were used to compute the average spectrum for each night.

## 4. COMPOSITIONAL ANALYSIS OF (35107) 1991 VH

The NIR spectrum of (35107) 1991 VH is shown in Figure 1. The spectrum shows absorption bands centered at ~1 and ~2 μm indicative of the presence of the minerals olivine and pyroxene. We used the Bus-DeMeo taxonomy classification Web tool and found that the object is classified as an Sq-type (DeMeo et al. 2009) in agreement with previous classifications. Spectral band parameters including band centers and the band area ratio (BAR) were measured using the procedure described in Sanchez et al. (2020). Band centers are measured after dividing out the linear continuum, and correspond to the position of the minimum reflectance value obtained by fitting a polynomial over the absorption bands. Band areas are defined as the area between the linear continuum and the data curve and are used to calculate the BAR, which is given by the ratio of the area of Band II to that of Band I. A temperature correction derived by Sanchez et al. (2012) was applied to the BAR value in order to account for differences between the surface temperature of the asteroid and the room temperature at which spectral calibrations used for compositional analysis were obtained. The Band I and II center were found to be 1.025±0.005 and 2.000±0.005 μm, respectively. We calculated a temperature-corrected BAR of 0.47±0.05.

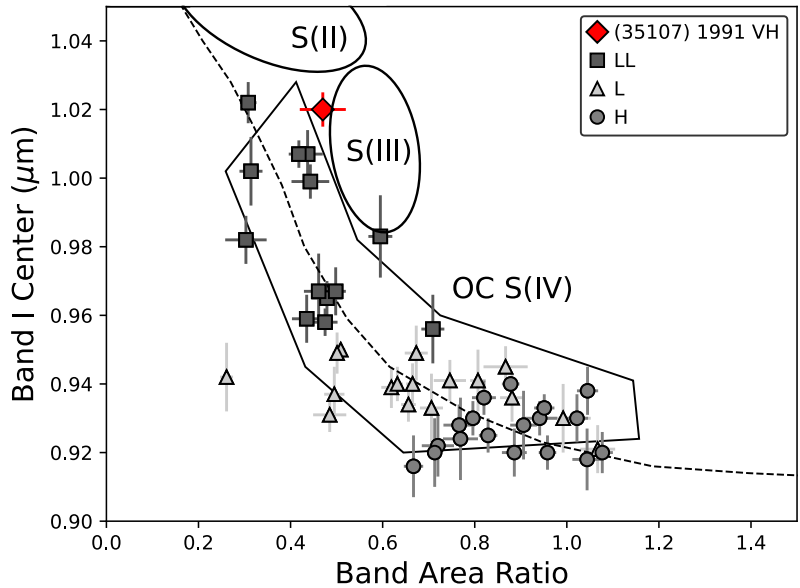

**Figure 3.** Band I center vs. BAR for (35107) 1991 VH. Values measured for LL, L and H ordinary chondrites from Sanchez et al. (2020) are depicted as squares, triangles and circles, respectively. Regions corresponding to the S(II), S(III) and S(IV) subgroups of Gaffey et al. (1993) are shown. OC

indicates the ordinary chondrites region, which is the same as the S(IV) subgroup region. The dashed line indicates the location of the olivine-orthopyroxene mixing line of Cloutis et al. (1986).

The Band I center vs. BAR of (35107) 1991 VH is shown in Figure 3. The asteroid falls in the upper part and slightly out of the S(IV) region corresponding to the ordinary chondrites as defined by Gaffey et al. (1993). The Band I center and BAR were used with the equations derived by Sanchez et al. (2020) to determine the composition of the asteroid. The Band I center is used to determine the olivine and pyroxene chemistries, which are given by the molar % of fayalite (Fa) and ferrosilite (Fs), respectively. The BAR, on the other hand, is used to calculate the olivine to pyroxene ratio (ol/(ol+px)). We obtained values of $Fa_{30.6\pm2.0}$ and $Fs_{25.3\pm1.4}$ for the olivine and pyroxene, respectively, and an ol/(ol+px) of 0.59±0.04. Figure 4 shows Fa vs. Fs for 1991 VH along with values measured for LL, L and H ordinary chondrite meteorites from Nakamura et al. (2011). As can be seen in this figure, the composition of 1991 VH is consistent with LL ordinary chondrites. This is also evident in Figure 5 where we plot Fa vs. ol/(ol+px), again the object falls within the region corresponding to LL chondrites.

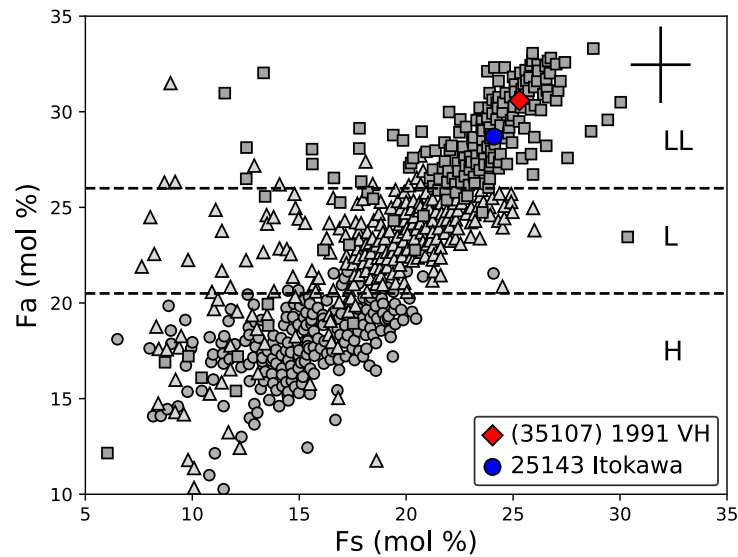

*Figure 4.* Mol % of fayalite (Fa) vs. ferrosilite (Fs) for (35107) 1991 VH. For comparison, calculated values for asteroid Itokawa from Sanchez et al. (2020) are also shown. Measured values for LL (squares), L (triangles), and H (circles) ordinary chondrites from Nakamura et al. (2011) are also included. The error bars in the upper right corner correspond to the uncertainties derived by Sanchez et al. (2020), 2.0 mol% for Fa, and 1.4 mol% for Fs. Figure adapted from Nakamura et al. (2011).

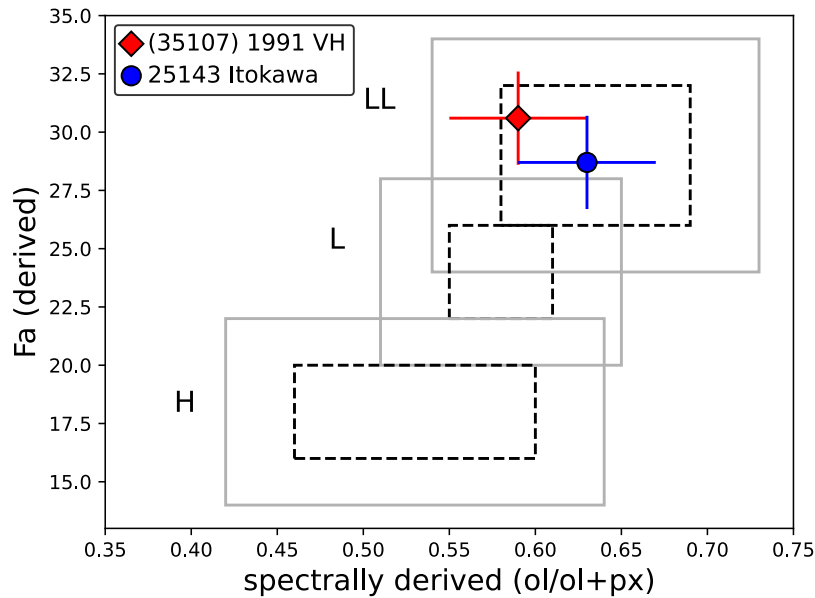

**Figure 5.** Molar content of Fa vs. ol/(ol+px) ratio for (35107) 1991 VH. Black dashed boxes represent the range of measured values for each ordinary chondrite subgroup. Gray solid boxes correspond to the uncertainties associated with the spectrally-derived values from Sanchez et al. (2020). Figure adapted from Dunn et al. (2010).

## 5. THERMAL MODELING AND PHASE REDDENING OF (175706) 1996 FG3

All the NIR spectra of (175706) 1996 FG3 show a thermal excess at wavelengths > 2 μm. This thermal flux was modeled following the same procedure described in Reddy et al. (2009, 2012). Thermal models were generated for a given solar distance and phase angle for albedos ranging from 1% to 10%. We assumed an emissivity $\varepsilon = 0.90$ and a beaming parameter $\eta=0.75$ for all the models. The spectra along with the modeled albedo curves that can be fit to encompass the observed thermal fluxes are shown in Figure 6. We found that the best fits correspond to albedos ranging from ~1.5 to 5%, although we noticed that for the spectra with the highest SNR, the best fits correspond to an albedo of ~2-3%. The models were then used to remove the thermal excess from each spectrum (Figure 6).

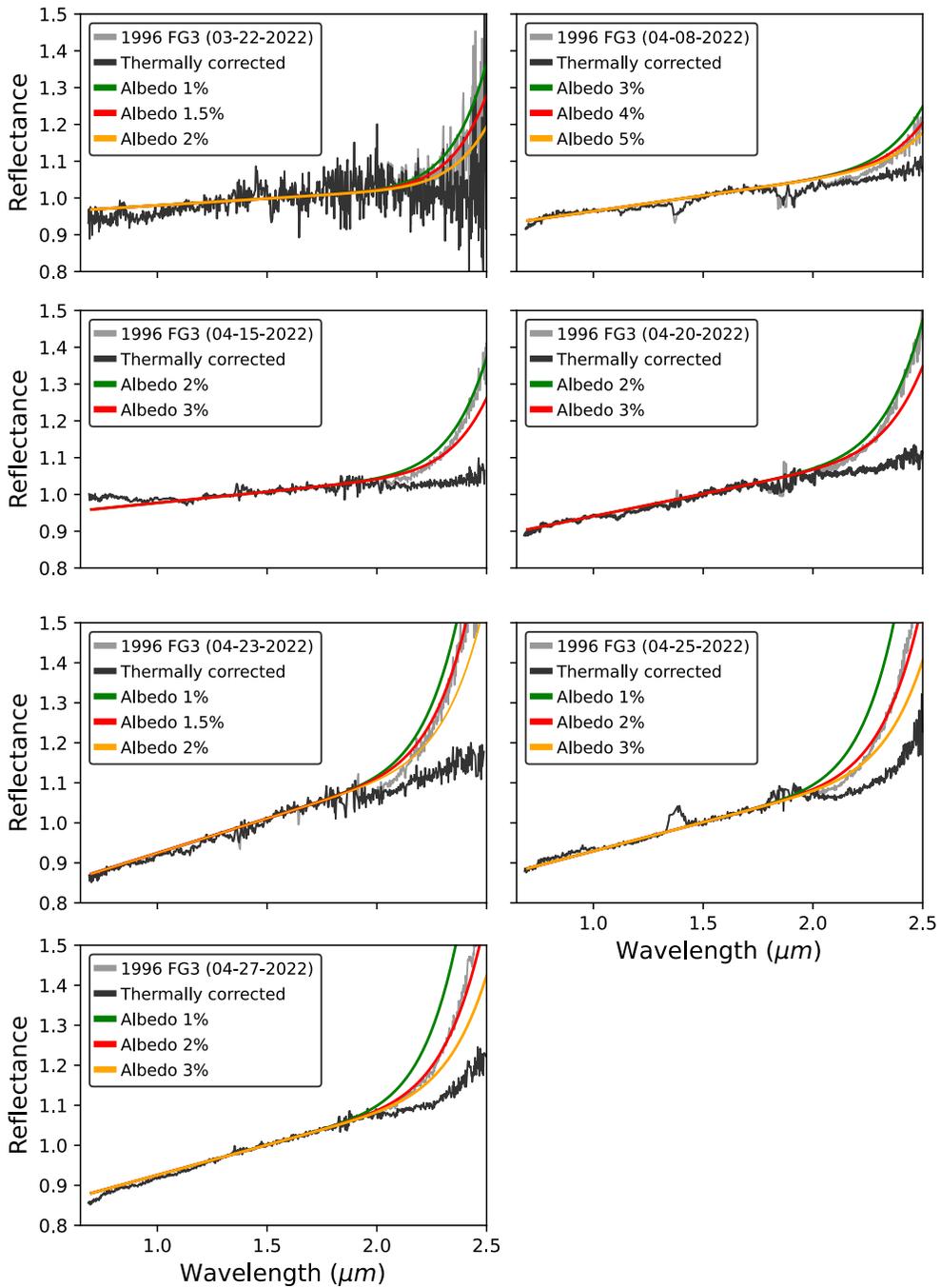

*Figure 6.* NIR spectra of (175706) 1996 FG3 normalized to unity at 1.5 µm. Each plot corresponds to the average spectrum for one night of observation. Modeled albedo curves that encompass the observed thermal fluxes and thermally corrected spectra are also shown.

Since the spectra were obtained at a broad range of phase angles (~13° to 53°), we looked for possible signs of phase reddening, which is known to produce an increase of the spectral slope as the phase angle increases (*e.g.*, Sanchez et al. 2012). For this, the spectral slope was calculated for each average spectrum from a linear fit to the data between 1.0 and 1.75 µm. Figure 7 shows the spectral slopes (measured in µm$^{-1}$) as a function of phase angle. For each measurement of the spectral slope the standard error was calculated. We found that the largest standard error is 0.006 and corresponds to the data obtained on 03-22-2022. However, it is worth noting that Marsset et al. (2020) found that variations in the spectral slope can result from the difference in air mass between the asteroid and the solar analog used. In particular, they reported a variation of -0.92%/microns per 0.1-unit air mass difference between the asteroid and the solar analog. Therefore, in order to account for this source of error, we calculated the difference between the air mass of the solar analog and the mean air mass of the asteroid for each night. The results were then divided by 0.1 and multiplied by 0.0092. We found that when the difference in air mass between the asteroid and the solar analog is considered, the uncertainty in the spectral slope ranges from ~0.018 to 0.034. These uncertainties were adopted as the errors for the spectral slope and are shown in Figure 7. Finally, in order to quantify the change in spectral slope with phase angle (*i.e.*, the degree of phase reddening), we performed a linear fit to the data and calculated the slope of the fit and the $R^2$.

Evidence of possible phase reddening can be seen for phase angles > 19°, with a continuous increase in the spectral slope as phase angle increases. Our results support the findings of de León et al. (2013) who reported an increase in spectral slope measured between 0.9 and 2.1 $µm$ for 1996 FG3 in the phase angle range of 6° to 51°. In their work this variation corresponds to 0.004±0.002 $µm^{-1}deg^{-1}$, which is in agreement with our slope value of 0.0027±0.001 $µm^{-1}deg^{-1}$ (Figure 7). This is also consistent with the monotonic phase dependence of the spectral slope observed for Bennu, a B-type asteroid, in the wavelength range 0.55-2.5 $µm$ (Fornasier et al. 2020). Moreover, the phase reddening slope we computed for 1996 FG3 (0.0027±0.001 $µm^{-1}deg^{-1}$) is close to the value computed for asteroid Ryugu (0.002±0.007 $µm^{-1}deg^{-1}$) using the Hayabusa2 color camera data from 0.40 to 0.95 $µm$ (Tatsumi et al. 2020). An even weaker phase reddening slope was inferred for Bennu using the OSISRIS-REx (Origins, Spectral Interpretations, Resource Identification, and Security—Regolith Explorer) visible and infrared spectrometer: 4.16±0.08 $10^{-4}µm^{-1}deg^{-1}$ for the 0.48-2.5 $µm$ range and 1.525±0.002 $10^{-3}µm^{-1}deg^{-1}$ for the 0.48-0.86 $µm$ range (Zou et al. 2020). Accordingly, it is possible that 1996 FG3 has similar physical properties (*e.g.*, microscopic roughness and porosity) as Bennu and Ryugu if these three asteroids share the same phase reddening effect (Tatsumi et al. 2020).

We checked for possible rotational variation of the spectrum of 1996 FG3 using the data from 04-27-2022 that includes five sets of spectra with very good SNR. After ratioing each set to the average for the night and normalizing at 1.5 microns, observed variations in the spectral slope and absorption features are less than ~3% outside the telluric bands. Data from other nights also support this finding. Therefore, 1996 FG3 appears to have a uniform mineralogical composition across its surface.

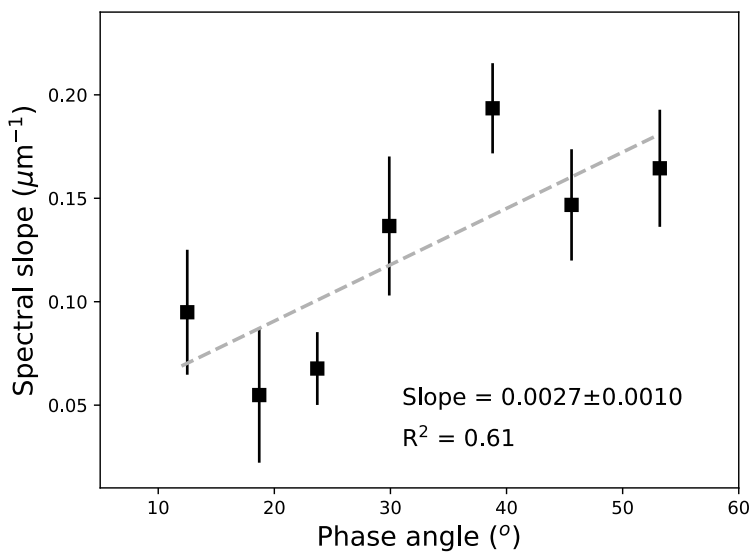

*Figure 7.* Spectral slopes calculated from a linear fit to the spectra between 1.0 and 1.75 μm vs. phase angle for (175706) 1996 FG3.

## 6. TAXONOMIC CLASSIFICATION AND METEORITE ANALOGS FOR 1996 FG3

For the taxonomic classification we combined the NIR spectrum of 1996 FG3 obtained on 04-15-2022 with the visible spectrum of Binzel et al. (2001). The NIR spectrum obtained on that date was chosen because it has the most neutral spectral slope and produces the best overlap with the visible spectrum. The Bus-DeMeo taxonomy classification Web tool yielded four possibilities: C-, Ch-, Xk-, or Xn-type. From a visual inspection we noticed that the Ch-type is the most similar (this taxonomic type also produced the smallest average absolute residual); however, it is worth mentioning that Ch-types are characterized as having an absorption band centered at ~0.7 µm, which is not visible in the spectrum of 1996 FG3 (Figure 8). The absence of this feature could be due to larger grain size (*e.g.*, Johnson and Fanale, 1973), or thermal metamorphism experienced by the body, which would make the 0.7 µm feature become weaker or disappear (*e.g.*, Cloutis et al. 2012).

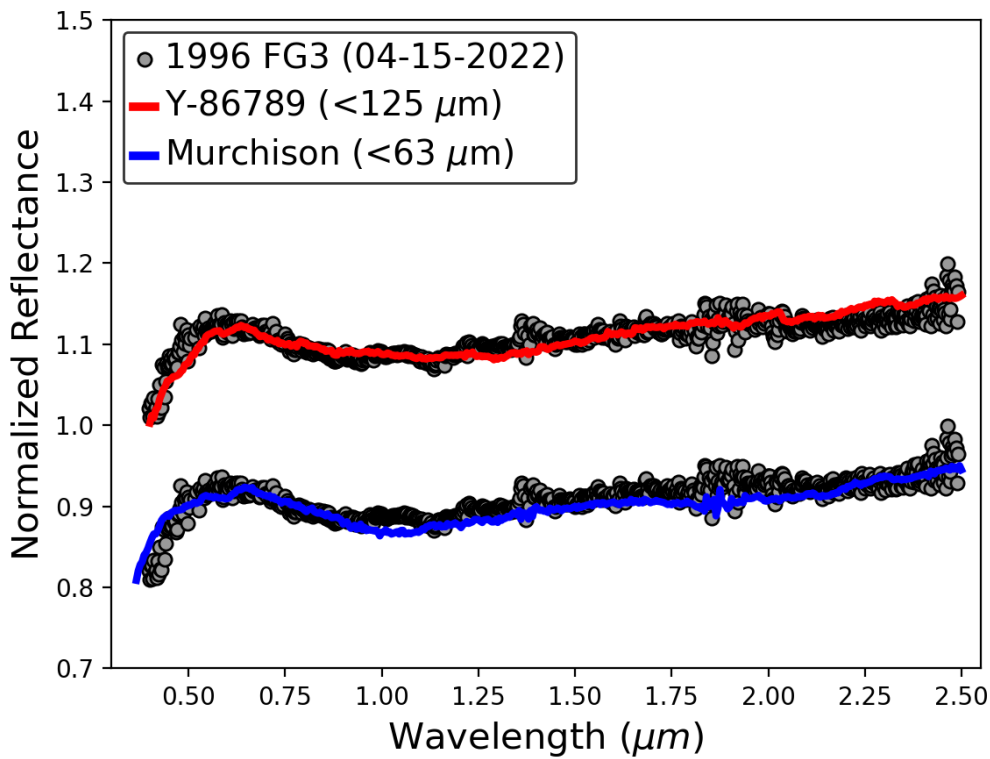

*Figure 8.* Visible and NIR spectrum of (175706) 1996 FG3. The NIR spectrum obtained on 04-15-2022 has been combined with the visible spectrum obtained by Binzel et al. (2001). Also shown are the spectrum of the C2$_{ung}$ carbonaceous chondrite Y-86789 (RELAB sample ID c1mp10) and the spectrum of the CM2 carbonaceous chondrite Murchison heated at 1000 °C (RELAB sample ID ncmb64g). The grain sizes of the meteorite samples are indicated. The spectra have been offset for clarity.

In order to look for possible meteorite analogs we used the Modeling for Asteroids (M4AST) online tool (Popescu et al. 2012), which allowed us to compare the spectrum of the asteroid with meteorite spectra from the RELAB database. The best spectral matches were obtained for two carbonaceous chondrites, Y-86789 and Murchison (Figure 8). Y-86789 is a C2$_{ung}$ chondrite that experienced substantial aqueous alteration and thermal metamorphism at temperatures of >500 °C (Matsuoka et al. 1996, King et al. 2019). Interestingly, the sample of the CM2 Murchison that produced the best match was heated for one week at 1000 °C. Previous studies have also identified CM carbonaceous chondrites as good meteorite analogs for (175706) 1996 FG3 (*e.g.*, de León et al. 2011).

## 7. SUMMARY

We observed two binary NEAs, the targets of NASA's Janus mission, using the 3-m NASA Infrared Telescope Facility on Mauna Kea, Hawaii:
1) 1991 VH was observed on July 26, 2008, when the asteroid had a visual magnitude of 15.1 and a phase angle of 85.7°,
2) 1996 FG3 was observed between March and April 2022, when the asteroid had visual magnitudes ranging from 15 to 17.1, and phase angles ranging from 12.5 to 53.2°.

Analysis of the resulting NIR spectra provided information on the asteroids' surface composition and meteorite analogs (LL ordinary chondrites for 1991 VH and thermally altered carbonaceous chondrites for 1996 FG3). In agreement with previous ground-based observations of 1996 FG3, we found an increase in spectral slope in the NIR when the phase angle increases (phase reddening effect). In addition, the phase reddening coefficient computed from a linear fit to the data ($0.0027\pm0.001$ $\mu m^{-1}deg^{-1}$) is similar to the phase reddening observed by space-based instruments on other carbonaceous asteroids, such as Ryugu and Bennu. No surface heterogeneity could be detected in our observations of 1996 FG3, as no rotational spectral variations were observed. These results could be confirmed using images acquired by the Janus spacecraft, providing an independent validation of Earth-based remote characterization methods.

## 8. ACKNOWLEDGMENTS

This material is based upon work supported by NASA under contract 80MSFC19C0038 issued through the Small, Innovative Missions for PLanetary Exploration (SIMPLEX) Program. This research work was also supported by NASA Near-Earth Object Observations grant 80NSSC20K0632. Data acquisition and processing was done in Tucson, Az. which is on the land and territories of Indigenous Peoples. We acknowledge our presence on the ancestral lands of the Tohono O'odham Nation and the Pascua Yaqui Tribe who have stewarded this area since time immemorial. This research utilizes spectra acquired with the NASA RELAB facility at Brown

University (http://www.planetary.brown.edu/relab/). The IRTF is operated by the University of Hawaii under contract 80HQTR19D0030 with NASA.